\documentclass[a4paper,fleqn,usenatbib]{mnras}

\usepackage[T1]{fontenc}
\usepackage{ae,aecompl}

\usepackage{graphicx}
\usepackage{amsmath}
\usepackage{amssymb}

\title[Helium settling in F stars]{Helium settling in F stars: constraining turbulent mixing using observed helium glitch signature}

\author[Verma et al.]{Kuldeep Verma$^{1}$\thanks{E-mail: kuldeep@phys.au.dk (KV)}
 and V\'{i}ctor Silva Aguirre$^{1}$\thanks{E-mail: victor@phys.au.dk (VSA)}
\\
$^{1}$Stellar Astrophysics Centre, Department of Physics and Astronomy, Aarhus University, Ny Munkegade 120, DK-8000 Aarhus C, Denmark
}

\date{Accepted XXX. Received YYY; in original form ZZZ}
\pubyear{2019}

\begin{document}
\label{firstpage}
\pagerange{\pageref{firstpage}--\pageref{lastpage}}
\maketitle

\begin{abstract}
Recent developments in asteroseismology -- thanks to space-based missions such as {\it CoRoT} and {\it Kepler} -- provide handles on those 
properties of stars that were either completely inaccessible in the past or only poorly measured. Among several such properties is the surface helium
abundance of F and G stars. We used the oscillatory signature introduced by the ionization of helium in the observed oscillation frequencies to 
constrain the amount of helium settling in F stars. For this purpose, we identified three promising F stars for which the standard models of atomic 
diffusion predict large settling (or complete depletion) of surface helium. Assuming turbulence at the base of envelope convection zone slows down 
settling of the helium and heavy elements, we found an envelope mixed mass of approximately $5 \times 10^{-4}$M$_\odot$ necessary to reproduce the 
observed amplitude of helium signature for all the three stars. This is much larger than the mixed mass of the order of $10^{-6}$M$_\odot$ found in 
the previous studies performed using the measurements of the heavy element abundances. This demonstrates the potential of using the helium signature 
together with measurements of the heavy element abundances to identify the most important physical processes competing against atomic diffusion, 
allowing eventually to correctly interpret the observed surface abundances of hot stars, consistent use of atomic diffusion in modelling both hot 
and cool stars, and shed some light on the long-standing cosmological lithium problem. 
\end{abstract}


\section{Introduction}
%---------------------
\label{intro}
Atomic diffusion is a fundamental physical process driven by pressure, temperature and composition gradients. The possibility of atomic diffusion 
in stellar interiors was first realized about a century ago \citep{chap17a,chap17b,chap22}. A significant impact of atomic diffusion on the solar 
surface heavy element abundances was first predicted by \citet{alle60}, which was later confirmed by the developments in helioseismology 
\citep[see e.g.][]{jcd93b,rich96,bahc97}. Moreover, the observed surface abundances of stars at different evolutionary stages in clusters can not be 
explained without atomic diffusion \citep[see e.g.][]{korn06,korn07,lind08,nord12,gruy13,gruy14,mott18}. 

An apparent problem with models of atomic diffusion arises when we consider stars more massive than the Sun, for which the envelope 
convection zone is confined very close to the surface, coinciding with the region of large pressure and temperature gradients. For such stars, 
models of atomic diffusion predict large settling (or complete depletion) of surface helium and heavy elements \citep[see e.g.][]{more02}, in 
contrast to the recent measurements of helium glitch signature in the observed oscillation frequencies of F stars \citep{verm17} and observations of 
the surface heavy element abundances of A and F stars \citep[see e.g.][]{vare99}. It is well known that radiative forces play an important role in 
such stars \citep[see e.g.][]{turc98,dott17,deal18}, however even after including them, the predicted surface abundance anomalies are still much 
larger than the observations \citep[see e.g.][]{turc98}. Particularly, radiative force on helium is negligible and can not slow down its settling, 
pointing towards the presence of other physical processes competing against atomic diffusion. 

The two best-studied physical processes that can effectively reduce the efficiency of atomic diffusion are: (1) turbulence at the base of envelope 
convection zone \citep[see e.g.][]{scha69,vauc78a,vauc78b}, and (2) mass loss from the surface \citep[see e.g.][]{mich83,mich86}. \citet{rich00} 
assumed turbulence and radiative forces as competing processes against atomic diffusion with a phenomenological model of turbulent diffusion to 
reproduce the observed heavy element abundances of AmFm stars. They found that the models do not reproduce the observations perfectly. However, 
given large systematic uncertainties associated with the observations, they argued it to be premature to conclude that hydrodynamical processes 
other than turbulence are needed. \citet{vick10} assumed mass loss and radiative forces as competing processes to reproduce the observed surface 
abundance anomalies of AmFm stars, and concluded that the current observational constraints are not sufficient to discriminate between turbulence and 
mass loss. \citet{mich11b} explored turbulence and mass loss separately as competing processes for the observations of Sirius A, finding similar 
conclusions as \citet{vick10}. The above studies were all carried out using measurements of the surface heavy element abundances. Recent advances
in stellar seismology have provided unprecedented constraints on properties of solar-like oscillators. In this study, we shall illustrate using 
the example of turbulence, for the first time, how we can use direct seismic constraints on the surface helium abundance of F stars to study 
the processes competing against atomic diffusion. 

The ionization of helium introduces a glitch in the acoustic structure of solar-type stars, and leaves an oscillatory signature in the observed 
oscillation frequencies, $\nu$ \citep[see e.g.][]{goug88a,voro88,goug90a}. The strength (or amplitude) of the oscillatory signature depends on the 
amount of helium present its ionization zone -- the larger the helium abundance, the larger the amplitude \citep[see e.g.][]{basu04,houd04,mont05}. 
The observed amplitude of helium signature has recently been used to infer the surface helium abundance of solar-type stars 
\citep{verm14a,gai18,verm19}. The observed large amplitude of helium signature of F stars can not be reproduced by the standard models of atomic 
diffusion because of their low (or zero) prediction of the surface helium abundance, and will be used in this study to constrain the amount of 
turbulent mixing necessary.

The paper is organized in the following order. We describe the target selection in Section~\ref{tar} and outline the method to extract the helium 
glitch signature from the oscillation frequencies in Section~\ref{amp}. The details of the set of stellar models used are presented in 
Section~\ref{mod}. The results are discussed in Section~\ref{results} and conclusions are summarized in Section~\ref{conc}.

\section{Target selection}
\label{tar} 
It is well known that models with higher masses show larger helium and heavy element settling. Furthermore, element settling also depends on the 
evolutionary state, partly because it takes time for elements to sink and partly due to the evolution of the thickness of the convective 
envelope. The low mass models start being fully convective on the pre-main sequence (PMS), and arrive on the zero age main-sequence (ZAMS) with a 
convective envelope. As models evolve along the main-sequence (MS), the envelope convection zone becomes shallower until a point close to the 
terminal age main-sequence (TAMS), when it begins to increase in depth. The settling of the helium and heavy elements follow closely the evolution 
of the thickness of the convective envelope, i.e. element settling is small at the beginning (close to the ZAMS), large during the middle of 
the MS, and small again at the end (close to the TAMS). 

The study of acoustic glitches requires high-quality seismic data because of small amplitudes of their signatures in the observed oscillation 
frequencies \citep[see e.g.][]{mazu14,verm17}. The {\it Kepler} asteroseismic LEGACY sample consisting of 66 main-sequence stars with the highest 
quality seismic data \citep{lund17,silv17} is ideal for such a study. We selected only those stars from the LEGACY sample for which the standard 
stellar models with atomic diffusion predict largest amount of surface helium settling. This selection criterion is motivated from the fact 
that, for such a star, the difference between the observed and model amplitude of the helium signature is anticipated to be the largest because of 
the largest difference between the surface helium abundance of the star and the model, providing the tightest possible constraint on the physical 
processes acting against atomic diffusion. 

\begin{table}
\centering
\scriptsize
\caption{Sample of stars studied in this work. The parameters are inferred/taken from \citet{silv17}.}
\label{tab1}
\begin{tabular}{cccccccc}
\hline\hline
  KIC  &  $M$ (M$_\odot$)  &  $X_c$  &  $T_{\rm eff}$ (K)  &  $[{\rm Fe}/{\rm H}]_s$ (dex)\\ 
\hline
  2837475  &  $[1.39,1.46]$  &  $[0.35,0.44]$  &  $6614\pm77$  &  $0.01\pm0.10$\\ 
  9139163  &  $[1.34,1.42]$  &  $[0.32,0.48]$  &  $6400\pm84$  &  $0.15\pm0.09$\\
 11253226  &  $[1.32,1.46]$  &  $[0.34,0.42]$  &  $6642\pm77$  &  $-0.08\pm0.10$\\
\hline
\end{tabular}
\end{table}

\begin{figure}
\includegraphics[width=\columnwidth]{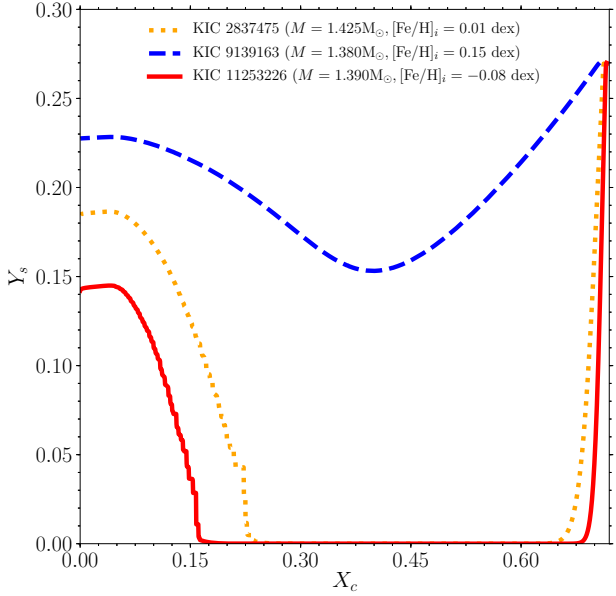}
\caption{Surface helium abundance as a function of central hydrogen abundance for the tracks representing KIC 2837475, 9139163 and 11253226. The 
tracks were computed with atomic diffusion of \citet{thou94}. The mass and initial metallicity of each track are shown in the legend. All tracks were 
computed with the initial helium abundance, mixing-length and exponential overshoot of $0.27$, $1.80$ and $0.016$, respectively.}
\label{fig1}
\end{figure}

We selected the appropriate targets based on their mass, $M$, and evolutionary state (central hydrogen abundance, $X_c$). The conservative ranges 
of $M$ and $X_c$ for stars in the LEGACY sample were obtained by taking the corresponding minimum and maximum values provided by the seven different 
fitting pipelines presented in \citet{silv17}. The targets were selected with lower limit on $M$ being greater than 1.3$M_\odot$ and $X_c$ being in 
the range $[0.3,0.5]$. This selection criterion left us with 3 stars: KIC 2837475, 9139163 and 11253226. They are all F stars listed in 
Table~\ref{tab1} along with their estimated $M$ and $X_c$ ranges and the observed effective temperature, $T_{\rm eff}$, and surface metallicity, 
$[{\rm Fe}/{\rm H}]_s$. 

To demonstrate that the representative models of these stars with atomic diffusion predict large settling of the surface helium, we computed three 
tracks with the mass and metallicity from Table~\ref{tab1} (central value of the range for the mass). Figure~\ref{fig1} shows the evolution (from 
right-to-left) of the surface helium abundance, $Y_s$, for the tracks. We can see the complete depletion of $Y_s$ for KIC 2839163 and 11253226 in 
the corresponding $X_c$ ranges listed in Table~\ref{tab1}, making them ideal for this study. The track corresponding to KIC 9139163 does not show 
complete depletion of surface helium because of its large metallicity (and hence thick convective envelope), however as we shall see in 
Section~\ref{results}, predicted settling is still too large for this star to reproduce the observed amplitude of helium signature.

\section{Average amplitude of helium signature}
%----------------------------------------------
\label{amp}
There are two popular fitting approaches to extract the glitch signatures from the oscillation frequencies: (1) by fitting directly the oscillation 
frequencies \citep[see e.g.][]{mont94,mont98,mont00}, and (2) by fitting their second differences, $\delta^2\nu_{n,l} = \nu_{n-1,l} - 2\nu_{n,l} + 
\nu_{n+1,l}$, where $n$ and $l$ are the radial order and harmonic degree, respectively \citep[see e.g.][]{goug90a,basu94,basu04}. In this work, we 
used a variation of the first approach as described in Method A of \citet{verm17,verm19}. It should be noted that the systematic uncertainties on 
the helium glitch parameters associated with different choices of fitting methods are generally small for stars in the LEGACY sample \citep{verm19}. 
This is because of the high-quality seismic data available for these stars with large numbers of detected modes and precisely measured oscillation 
frequencies. The precision of the observed oscillation frequencies of F stars is poor in comparison to other stars in the LEGACY sample because 
of their large linewidths \citep[see e.g.][]{appo12,lund17,comp19}, however their large amplitude of the helium signature effectively compensates for the 
lower precision.

For the convenience of the reader, we outline here the steps to calculate the average amplitude of helium signature. The stellar oscillation 
frequencies were fitted to the function,
\begin{equation}
f(n, l) = \nu_{\rm smooth} + \delta\nu_{\rm He} + \delta\nu_{\rm CZ}.
\end{equation}
The first (also the dominant) term represents the contribution to the oscillation frequency from the smooth structure of the star, which is modelled 
using $l$-dependent fourth degree polynomial in $n$ \citep[see e.g.][]{verm19},
\begin{equation}
\nu_{\rm smooth} = \sum_{k = 0}^4 b_k(l) n^k,
\end{equation}
where $b_k(l)$ are the polynomial coefficients to be determined by fitting $f(n, l)$ to the oscillation frequencies. The second and third terms are 
small perturbations to frequencies due to the helium and base of convection zone glitches, respectively. The functional form for these contributions 
are adapted from \citet{houd07},
\begin{eqnarray}
\delta\nu_{\rm He} &=& A_{\rm He} \nu e^{-8\pi^2\Delta_{\rm He}^2\nu^2} \sin(4\pi\tau_{\rm He}\nu + \psi_{\rm He}),\label{he}\\
\delta\nu_{\rm CZ} &=& \frac{A_{\rm CZ}}{\nu^2} \sin(4\pi\tau_{\rm CZ}\nu + \psi_{\rm CZ}),\label{cz}
\end{eqnarray}
where the parameters $A_{\rm He}$, $\Delta_{\rm He}$, $\tau_{\rm He}$, $\psi_{\rm He}$, $A_{\rm CZ}$, $\tau_{\rm CZ}$ and $\psi_{\rm CZ}$ are free
parameters to be again determined by fitting $f(n, l)$ to the oscillation frequencies. We used Monte Carlo simulation to propagate the observational 
uncertainties on oscillation frequencies to the fitted parameters.

We used the fitted parameters, $A_{\rm He}$ and $\Delta_{\rm He}$, to compute the average amplitude of the helium signature \citep{verm19},
\begin{eqnarray}
\langle A_\nu \rangle &=& \frac{\int_{\nu_1}^{\nu_2} A_{\rm He} \nu e^{-8\pi^2\Delta_{\rm He}^2\nu^2} d\nu}{\int_{\nu_1}^{\nu_2} d\nu}\nonumber\\
&=& \frac{A_{\rm He} [e^{-8\pi^2\Delta_{\rm He}^2\nu_1^2} - e^{-8\pi^2\Delta_{\rm He}^2\nu_2^2}]}{16\pi^2\Delta_{\rm He}^2[\nu_2 - \nu_1]},
\label{amplitude}
\end{eqnarray}
where $\nu_1$ and $\nu_2$ are the smallest and largest observed frequencies used in the fit. The same values of $\nu_1$ and $\nu_2$ have been 
consistently used to calculate the corresponding model $\langle A_\nu \rangle$.

\section{Stellar models}
\label{mod}
We used the stellar evolution code Modules for Experiments in Stellar Astrophysics \citep[MESA;][]{paxt11,paxt13,paxt15} to compute several grids of models with atomic 
diffusion and turbulence as competing process. The code was used with Opacity Project (OP) high-temperature opacities \citep{badn05,seat05} 
supplemented with low-temperature opacities of \citet{ferg05}. The metallicity mixture from \citet{gs98} was used. We used OPAL equation of state 
\citep{roge02}. The reaction rates were from NACRE \citep{angu99} for all reactions except $^{14}{\rm N}(p,\gamma)^{15}{\rm O}$ and 
$^{12}{\rm C}(\alpha,\gamma)^{16}{\rm O}$, for which updated reaction rates from \citet{imbr05} and \citet{kunz02} were used. We included settling 
of the helium and heavy elements following \citet{thou94}. We used an exponential overshoot at all possible radiative-convective boundaries 
\citep{herw00}. The oscillation frequencies were calculated using the Adiabatic Pulsation code \citep[ADIPLS;][]{jcd08a}.

\begin{figure}
\includegraphics[width=\columnwidth]{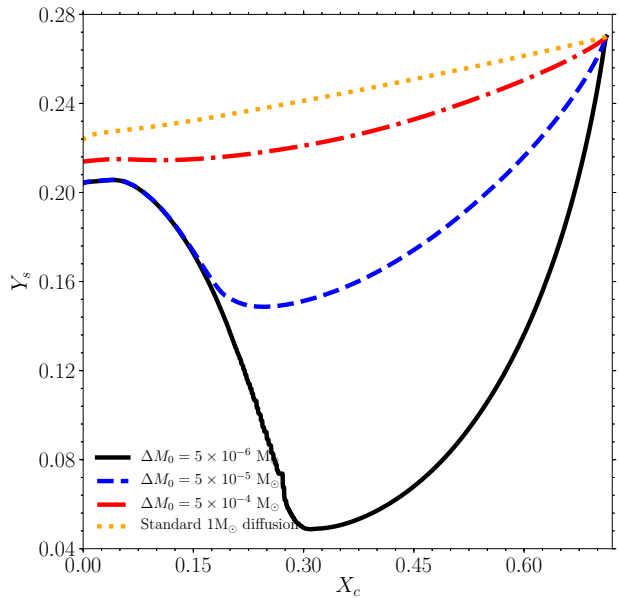}
\caption{Surface helium abundance as a function of central hydrogen abundance for tracks of masses 1.0 and 1.4M$_\odot$. The track with mass 
1M$_\odot$ (dotted curve) was computed with only atomic diffusion while those with mass 1.4M$_\odot$ (continuous, dashed and dot-dashed curves) were
computed with atomic diffusion and turbulence. The initial helium mass fraction, initial metal mass fraction, mixing-length and overshoot for each  
track were 0.27, 0.018, 1.8 and 0.015, respectively.}
\label{fig2}
\end{figure}

\subsection{Turbulent diffusion}
The stellar evolution code MESA provides a number of options for the input physics that the users can explore. For instance, it already implements a form of turbulent 
diffusion at the base of convection zone (see the module {\it \$MESA\_DIR/star/private/turbulent\_diffusion.f90} in the MESA package) to study the 
helium and heavy element settling in the Sun \citep{prof91}. Moreover, it also ships a number of hooks with the package to enable the users 
conveniently implement certain type of new physics. 

We used the already present turbulent diffusion for the Sun in MESA as a guide to implement the turbulent diffusion coefficient of \citet{rich00} 
for hotter stars. This was accomplished using the hook {\it \$MESA\_DIR/star/other/other\_d\_mix.f90}. We implemented the density-dependent 
diffusion coefficient,
\begin{equation}
D_{\rm T} = \omega D({\rm He})_0 \left(\frac{\rho_0}{\rho}\right)^n,
\label{turb}
\end{equation}
where $\omega$ and $n$ are constants. Here, $\rho_0$ and $D({\rm He})_0$ are the density and atomic diffusion coefficient of helium at a reference 
depth, respectively. We used a simple analytical approximation for the atomic diffusion coefficient of helium following \citet{rich00}, 
\begin{equation}
D({\rm He}) = \frac{3.3 \times 10^{-15} T^{2.5}}{4 \rho \ln(1 + 1.125 \times 10^{-16} T^3 /\rho)} \ ({\rm in \ cgs \ units}),
\end{equation}
where $T$ is the temperature. In earlier studies, reference depth in Eq.~\ref{turb} was defined at either fixed $\rho$ or $T$ 
\citep[see e.g.][]{rich00}, however the definition has been revised in recent studies as follows. \citet{rich00} 
noted that the surface element abundances depend only on the envelope mass mixed by turbulence (and not on the independent choices of $\omega$, $n$ 
and reference depth). This led \citet{mich11a,mich11b} to cleverly redefine $D_{\rm T}$ by fixing $\omega$ at 10000 and $n$ at 4 and anchoring 
the turbulent diffusion coefficient at a radial coordinate, $r_0$, determined by a fixed outer mass, $\Delta M_0 = M - M_0$, where $M_0$ is the mass 
enclosed in a sphere of radius $r_0$. This means that $\rho_0$ and $D({\rm He})_0$ in Eq.~\ref{turb} are the density and atomic diffusion coefficient 
of helium at a depth corresponding to the outer mass of $\Delta M_0$. Since the parameter $\Delta M_0$ determines the envelope mass mixed by turbulence 
\citep[see e.g.][]{mich11b}, it is referred as envelope mixed mass in the subsequent sections. Note that the envelope convection zone of F 
stars may be split into superficial convective regions due to opacity peaks in the ionization zones of helium and iron. The turbulent mixing was 
used at the base of every such region, resulting complete mixing of radiative layers between superficial convection zones, as expected from the 
results of numerical simulations \citep[see e.g.][]{kupk02,frey04}. 

Figure~\ref{fig2} shows evolution of the surface helium abundance for models in three tracks with $\Delta M_0 = 5 \times 10^{-6}, 
5 \times 10^{-5} \ {\rm and} \ 5 \times 10^{-4}$M$_\odot$. All three tracks were computed with mass $1.4$M$_\odot$. We can see that as $\Delta M_0$ 
increases, the amount of helium settling on the MS decreases. The models of atomic diffusion have been thoroughly tested for the Sun using 
helioseismic data, and are known to predict helium and heavy element settling reasonably well without any additional mixing. Figure~\ref{fig2} also 
shows a solar-mass track without turbulent mixing for comparison. We note that the track with $\Delta M_0 = 5 \times 10^{-4}$M$_\odot$ is closest to 
the solar-mass track. 

\begin{table*}
\centering
\scriptsize
\caption{Parameter spaces considered for each grid of stars in the sample. The range in initial metallicity of a grid depends on the envelope mixed 
mass (see the text).}
\label{tab2}
\begin{tabular}{cccccccc}
\hline\hline
  KIC  &  $M$ (M$_\odot$)  &  \multicolumn{3}{c}{$[{\rm Fe}/{\rm H}]_i$}  &  $Y_i$  &  $\alpha_{\rm MLT}$  &  $f_{\rm OV}$\\
\cline{3-5}\\
       &                   &  $\Delta M_0 = 5 \times 10^{-6}$M$_\odot$  &  $\Delta M_0 = 5 \times 10^{-5}$M$_\odot$  &  $\Delta M_0 = 5 \times 10^{-4}$M$_\odot$ &  &  &\\
\hline
  2837475  &  $[1.39,1.46]$  &  $[0.12,0.32]$  &  $[0.12,0.32]$  &  $[0.00,0.20]$  &  $[0.24,0.30]$  &  $[1.5,2.1]$  &  $[0.00,0.03]$\\ 
  9139163  &  $[1.34,1.42]$  &  $[0.15,0.35]$  &  $[0.15,0.35]$  &  $[0.12,0.32]$  &  $[0.24,0.30]$  &  $[1.5,2.1]$  &  $[0.00,0.03]$\\
 11253226  &  $[1.32,1.46]$  &  $[0.07,0.27]$  &  $[0.05,0.25]$  &  $[-0.08,0.12]$ &  $[0.24,0.30]$  &  $[1.5,2.1]$  &  $[0.00,0.03]$\\
\hline
\end{tabular}
\end{table*}

\begin{figure*}
\includegraphics[width=\textwidth]{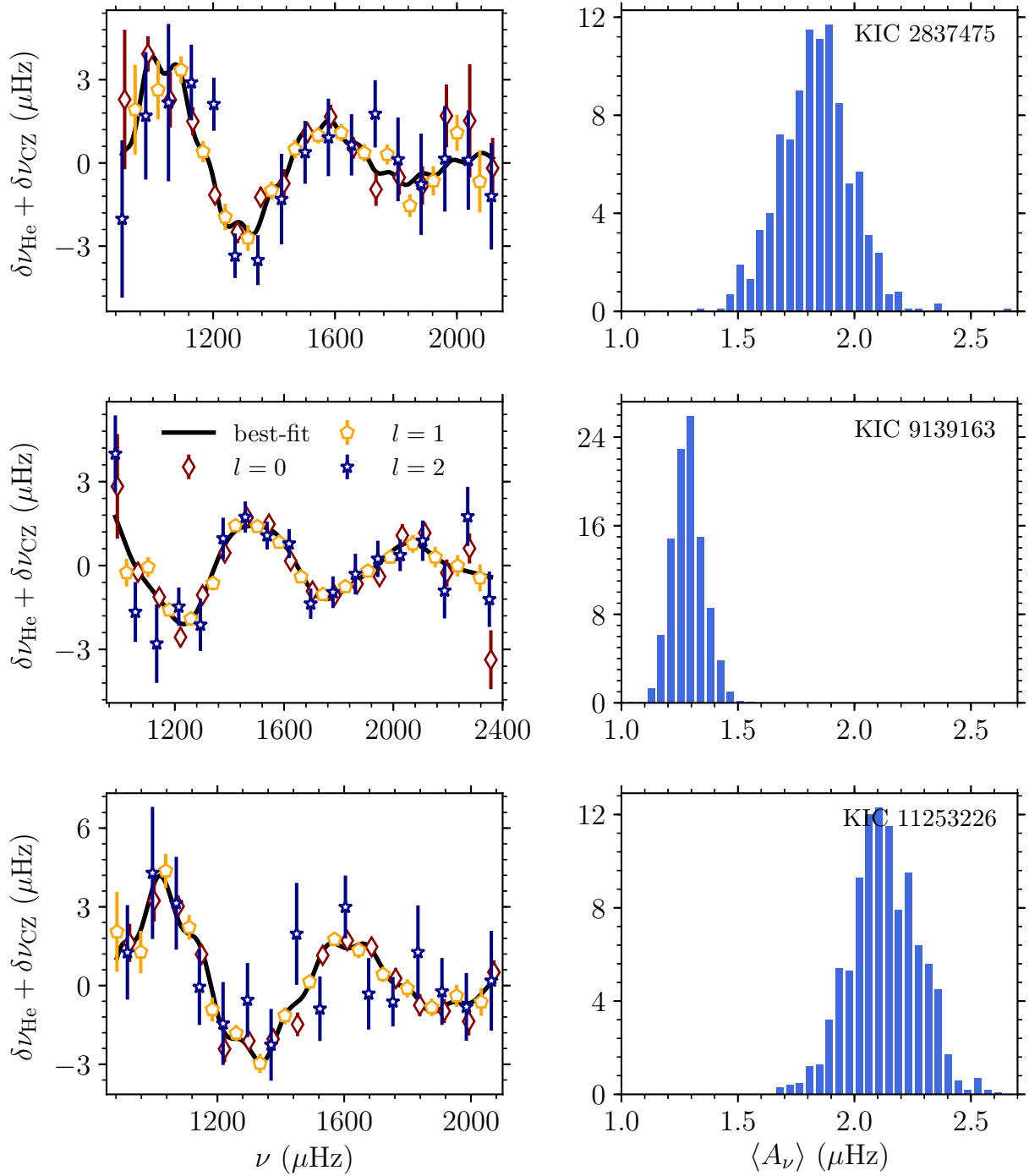}
\caption{Fit to the observed oscillation frequencies of KIC 2837475, 9139163 and 11253226 (smooth component has been subtracted to clearly see the
glitch signatures). The different rows correspond to the three stars. In the left panels, the different types of points show the observed 
modes of harmonic degrees 0, 1, and 2 while the curve represents the best-fit to them. In the right panels, the histograms show the distribution of 
average amplitude of the helium signature obtained using Monte Carlo simulation.}
\label{fig3}
\end{figure*}

It is clear from Eq.~\ref{turb} that $D_{\rm T}$ decreases very rapidly as we move inward from the anchor point because of the increase in density. 
This means that if the anchor point falls well within the envelope convection zone, i.e. $\Delta M_0 << \Delta M_{\rm CZ}$, then the turbulent 
mixing has very little impact at the base of convection zone. Since the low-mass models have thick convective envelope, the inclusion of 
turbulence has insignificant impact on their structure (as desired for the Sun). The evolution of the surface helium abundance for the solar-mass
track with atomic diffusion and turbulence falls exactly on the dotted curve in Figure~\ref{fig2} (not shown for clarity). This opens a possibility 
for the future to consistently use atomic diffusion with turbulence while modelling both low and high-mass stars, reducing the systematic 
uncertainties on the inferred stellar properties associated with the arbitrary transition from diffusion models for low-mass stars to non-diffusion 
models for high-mass stars.

\subsection{Model grids}
We constructed three grids of models for each star with three values of the envelope mixed mass, $\Delta M_0 = 5 \times 10^{-6}, 
5 \times 10^{-5} \ {\rm and} \ 5 \times 10^{-4}$M$_\odot$. For each grid, we computed 50 tracks -- each containing hundreds of models on the 
MS -- sampling uniformly using quasi-random numbers (more specifically using Sobol sequences) the 5-D space formed by mass $M$, initial metallicity 
$[{\rm Fe}/{\rm H}]_i$, initial helium abundance $Y_i$, mixing-length $\alpha_{\rm MLT}$ and overshoot $f_{\rm OV}$. Note that we do not need very 
dense grid of models like we do when modelling stars. The goal here is to explore the likely parameter space of stars in the sample to see if we 
can reproduce the observed amplitude of helium signature. Table~\ref{tab2} lists the parameter spaces considered for each grid of stars in the 
sample. The range in $[{\rm Fe}/{\rm H}]_i$ was shifted to higher values in comparison to the observed $[{\rm Fe}/{\rm H}]_s$ to compensate for 
element settling. Note that the shift depends on the value of $\Delta M_0$ and was estimated by trying different values of $[{\rm Fe}/{\rm H}]_i$ 
and comparing the predicted $[{\rm Fe}/{\rm H}]_s$ with the corresponding observed value for each $\Delta M_0$. Since smaller value of 
$\Delta M_0$ means larger settling of the helium and heavy elements on the MS, the shift is larger for the grid with 
$\Delta M_0 = 5 \times 10^{-6}$M$_\odot$ in comparison to $\Delta M_0 = 5 \times 10^{-4}$M$_\odot$. 

We find a representative model of a star from a given track by fitting the surface corrected model frequencies \citep{kjel08} to the observed ones.
As discussed in \citet{verm19}, the choice of the surface correction scheme is not important for this particular exercise because, for a given
track, the age can be determined very precisely by fitting the observed oscillation frequencies irrespective of the surface correction used. In this
manner, we get 50 representative models of a star from each grid.

\section{Results}
%----------------
\label{results}
We fitted the observed oscillation frequencies of each star using the method outlined in Section~\ref{amp} to extract the glitch signatures. 
The left panels in Figure~\ref{fig3} show the fit to the observed oscillation frequencies of KIC 2837475, 9139163 and 11253226 (top-to-bottom) after 
removing the smooth component, $\nu_{\rm smooth}$. The smooth component was subtracted to clearly see the glitch signatures. The signature with 
large amplitude and large period is from the helium ionization zone while the modulation on top of it with small amplitude and small period is from 
the base of convection zone. The right panels in the figure show the distribution of the observed average amplitude of the helium glitch 
signature obtained using Monte Carlo simulation. The unimodal nature of the distribution shows the robustness of the fit. As discussed in 
Section~\ref{amp}, we can see the large average amplitude of the helium signature (>1$\mu$Hz) for these stars in comparison to the other cool stars 
in the LEGACY sample (<1$\mu$Hz; see Table~1 of \citet{verm17}).

As discussed in Section~\ref{mod}, we have three sets of 50 representative models of each star with three different values of the envelope mixed 
mass, $\Delta M_0 = 5 \times 10^{-6}, 5 \times 10^{-5} \ {\rm and} \ 5 \times 10^{-4}$M$_\odot$. We fitted the frequencies of all the representative 
models to extract the glitch signatures. To avoid systematic uncertainties, the fit was performed using the same set of modes and weights as for 
the observations. We used the fitted parameters for all the models to calculate the corresponding average amplitude of the helium signature using 
Eq.~\ref{amplitude}. 

\begin{figure}
\includegraphics[width=\columnwidth]{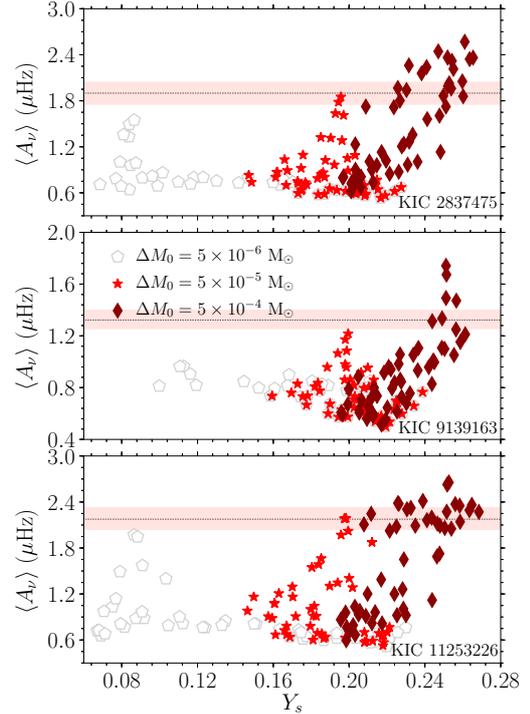}
\caption{Average amplitude of the helium glitch signature as a function of the surface helium abundance. The different panels correspond to the 
three stars. The different types of points in a panel represent the three different sets of 50 representative models with different values of the 
envelope mixed mass. The horizontal dotted line corresponds to the observed amplitude with the band representing 1$\sigma$ uncertainty.}
\label{fig4}
\end{figure}

\begin{figure}
\includegraphics[width=\columnwidth]{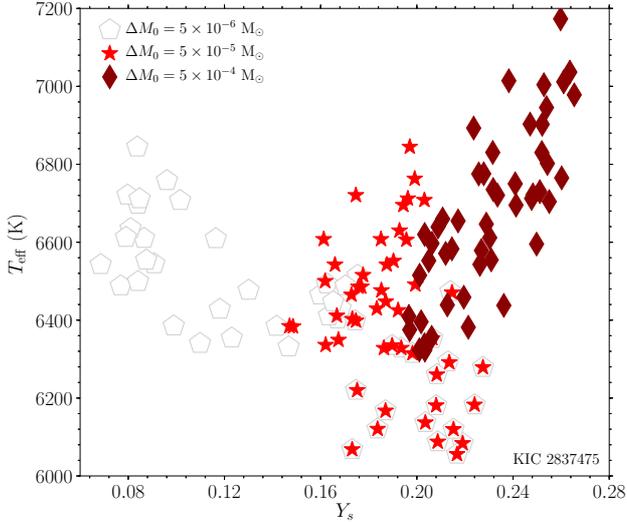}
\caption{Effective temperature as a function of the surface helium abundance for the models of KIC 2837475. The different types of points in a 
panel represent the three different sets of 50 representative models with different values of the envelope mixed mass.}
\label{fig5}
\end{figure}

\subsection{Comparison of the observed and model amplitudes}
%-----------------------------------------------------------
We compare in Figure~\ref{fig4} the observed average amplitude of the helium signature with those predicted by the representative models of different 
envelope mixed mass. First, we shall look at the models with $\Delta M_0 = 5 \times 10^{-6}$M$_\odot$. In all three stars it can be seen that $Y_s$ 
for most of the representative models is small because of the large helium settling. The small $Y_s$ introduces a weak helium glitch in the acoustic 
structure, leaving a weak helium signature in the model frequencies. Consequently, the average amplitude is smaller for all models in comparison to 
the observed $\langle A_\nu \rangle$. This led us to conclude that $\Delta M_0$ must be greater than $5 \times 10^{-6}$M$_\odot$ to reduce further 
the helium settling. 

In Figure~\ref{fig4}, it is interesting to note the behaviour of $\langle A_\nu \rangle$ as a function of $Y_s$ for the models with 
$\Delta M_0 = 5 \times 10^{-6}$M$_\odot$; it decreases slightly instead of increasing as $Y_s$ increases. This is because, it turns out that the 
models with smaller $Y_s$ have larger $T_{\rm eff}$ for $\Delta M_0 = 5 \times 10^{-6}$M$_\odot$, as can be seen in Figure~\ref{fig5}. The average 
amplitude is known to rapidly increase as $T_{\rm eff}$ increases \citep[see Figure~6 of][]{verm14b}. Hence, the effect of the decrease in $Y_s$ 
on $\langle A_\nu \rangle$ gets compensated by the increase in $T_{\rm eff}$, resulting in the trend seen in the figure.
 
As we can see in Figure~\ref{fig4}, most of the models (all for KIC 9139163) with $\Delta M_0 = 5 \times 10^{-5}$M$_\odot$ have systematically lower
average amplitude than the corresponding observed $\langle A_\nu \rangle$. This suggests that these models are also unlikely to represent the 
stars, though we can not completely disregard the possibility yet. The models with $\Delta M_0 = 5 \times 10^{-4}$M$_\odot$, however, show the 
familiar trend, i.e. $\langle A_\nu \rangle$ increases as $Y_s$ increases. These models do reproduce the observed $\langle A_\nu \rangle$ for 
certain value of $Y_s$. 

We note that the scatter in Figure~\ref{fig4} for a given $\Delta M_0$ is intrinsic and due to differences in $M$, $[{\rm Fe}/{\rm H}]_i$, 
$\alpha_{\rm MLT}$, $f_{\rm OV}$ and age. Although $\langle A_\nu \rangle$ correlates with other stellar parameters such as age, the 
correlation can mostly be explained by the dependence of the corresponding parameter on $Y_s$ and $T_{\rm eff}$ \citep[see Figure~7 of][]{verm14a}. 
Moreover, we constructed grids with broad ranges in stellar parameters (for an example, models between the ZAMS and the TAMS were considered) 
in order to avoid any possible biases (for an example, related to the uncertainties in the evolutionary stage of the targets). 

\begin{figure}
\includegraphics[width=\columnwidth]{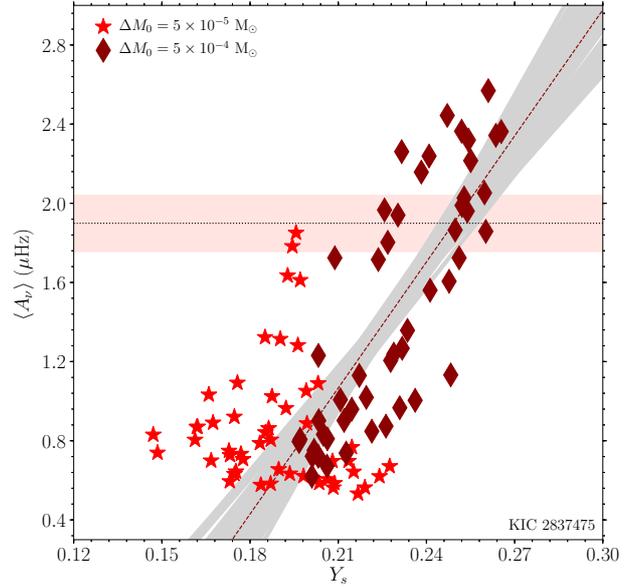}
\caption{Same as the panel for KIC 2837475 in Figure~\ref{fig4}, except the models with $\Delta M_0 = 5 \times 10^{-6}$M$_\odot$ are not shown. The 
dashed line is a straight line fit to the models with the band representing Monte Carlo regression uncertainty.}
\label{fig6}
\end{figure}

In Figure~\ref{fig6}, we note that the models with $\Delta M_0 = 5 \times 10^{-5} \ {\rm and} \ 5 \times 10^{-4}$M$_\odot$ put together show a 
reasonably tight correlation between $\langle A_\nu \rangle$ and $Y_s$. We calibrated the observed $\langle A_\nu \rangle$ against these models of 
different $Y_s$ to infer the surface helium abundance for all three stars. To perform the calibration, we fitted a straight line to the model 
$\langle A_\nu \rangle$ as a function of $Y_s$. Since the models with $\Delta M_0 = 5 \times 10^{-5}$M$_\odot$ are unlikely to represent the stars 
and may potentially bias the helium estimate, we gave less weight to these models in the fit. The trial and error method suggested a reasonable 
choice of one-fourth of the weight given to the models with $\Delta M_0 = 5 \times 10^{-4}$M$_\odot$. The intersection of the fitted line and the 
horizontal line corresponding to the observed $\langle A_\nu \rangle$ provided $Y_s$ of $0.249_{-0.007}^{0.007}$, $0.264_{-0.006}^{+0.007}$ and 
$0.254_{-0.006}^{+0.007}$ for KIC 2837475, 9139163 and 11253226, respectively.

\begin{figure}
\includegraphics[width=\columnwidth]{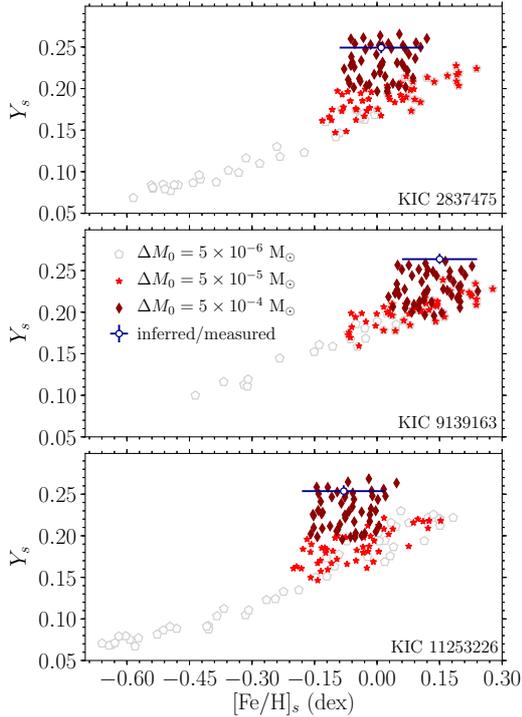}
\caption{Surface helium abundance as a function of the surface metallicity. The different panels correspond to the three stars. The different types 
of points without errorbar in a panel represent the three different sets of 50 representative models with different values of the envelope mixed 
mass. The point with errorbar represents the inferred/measured surface helium and metallicity of the star.}
\label{fig7}
\end{figure}

In Figure~\ref{fig7}, we compare simultaneously the inferred $Y_s$ and measured $[{\rm Fe}/{\rm H}]_s$ of all stars with the corresponding 
quantities predicted by the representative models with different values of $\Delta M_0$. We can see that only the models with $\Delta M_0 = 
5 \times 10^{-4}$M$_\odot$ reproduce both quantities together. This led us to conclude that we need turbulent mixing with approximately 
$\Delta M_0 = 5 \times 10^{-4}$M$_\odot$ to reproduce the observed $\langle A_\nu \rangle$.
 
\begin{figure}
\includegraphics[width=\columnwidth]{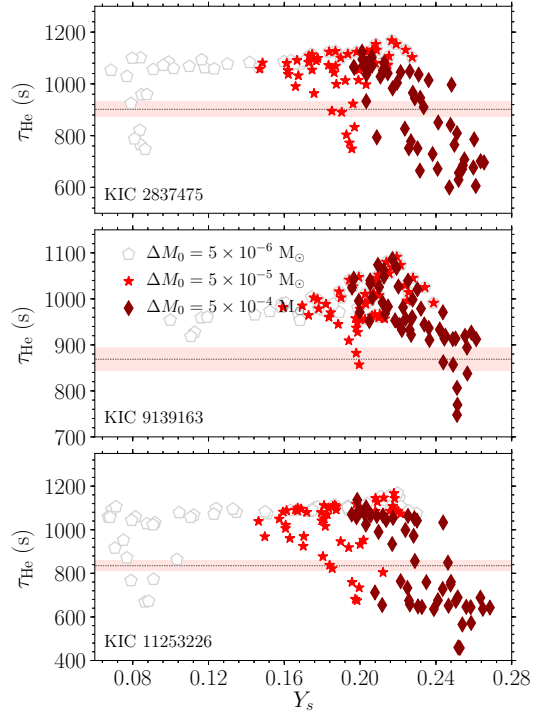}
\caption{Acoustic depth of the helium ionization zone as a function of the surface helium abundance. The different panels correspond to the 
three stars. The different types of points in a panel represent the three different sets of 50 representative models with different values of the 
envelope mixed mass. The horizontal dotted line corresponds to the observed acoustic depth with the band representing 1$\sigma$ uncertainty.
}
\label{fig8}
\end{figure}

\subsection{Comparison of the observed and model acoustic depths}
%----------------------------------------------------------------
We now compare the acoustic depth of the helium ionization zone obtained by fitting the observed oscillation frequencies with those obtained by 
fitting the model frequencies of different envelope mixed mass. As we can see in Figure~\ref{fig8} for all three stars, most of the models 
with $\Delta M_0 = 5 \times 10^{-6} \ {\rm and} \ 5 \times 10^{-5}$M$_\odot$ have systematically larger $\tau_{\rm He}$ than the corresponding 
observed $\tau_{\rm He}$. Since the acoustic depth of a layer depends on the sound speed of layers above it (see Eq.~\ref{tau}), most of these 
models have systematically different sound speed in the outer layers compared to the star. Again, models with 
$\Delta M_0 = 5 \times 10^{-4}$M$_\odot$ have $\tau_{\rm He}$ on both side of the observed $\tau_{\rm He}$. This reinforces the conclusion that we 
need turbulent mixing with approximately $\Delta M_0 = 5 \times 10^{-4}$M$_\odot$. 

\begin{figure}
\includegraphics[width=\columnwidth]{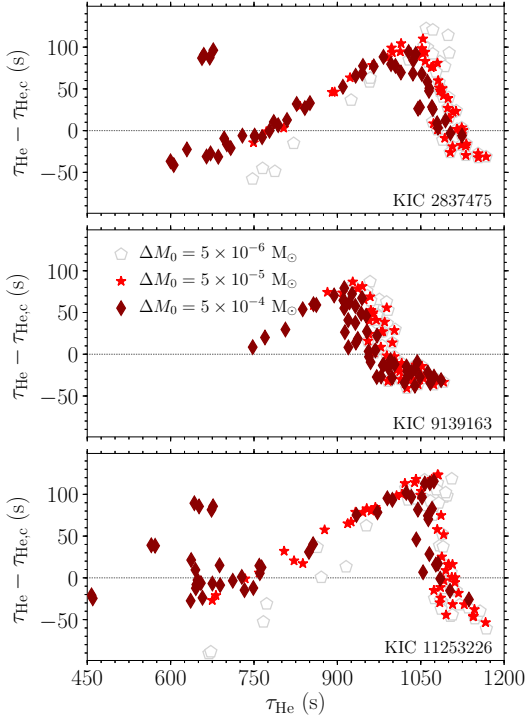}
\caption{Absolute difference between the two estimates of the acoustic depth of helium ionization zone as obtained by fitting the helium 
signature, $\tau_{\rm He}$, and by using the sound speed profile of the model, $\tau_{\rm He,c}$. The different panels correspond to the three 
stars. The different types of points in a panel represent the three different sets of 50 representative models with different values of the envelope 
mixed mass. The horizontal dotted line marks the zero difference.}
\label{fig9}
\end{figure}

\subsection{Sanity check for the model fits}
%-------------------------------------------
Fitting the oscillation frequencies to function $f(n, l)$ involves non-linear optimization in a high-dimensional space, and fits may not necessarily 
converge for all models. Since we have a large number of models, it is not feasible to examine the fits to the model frequencies visually as we do 
for the fits to the observed frequencies. We do, however, perform a sanity check by comparing the fitted acoustic depth of helium ionization zone 
($\tau_{\rm He}$ in Eq.~\ref{he}) with that obtained using sound speed, $c$, of the model. Using sound speed profile of the model, we can compute 
the acoustic depth of helium ionization zone, 
\begin{equation}
\tau_{\rm He,c} = \int_{R_{\rm He}}^{R_*} \frac{dr}{c},
\label{tau}
\end{equation} 
where $R_*$ and $R_{\rm He}$ are the radial coordinates of the acoustic surface and helium ionization zone, respectively. 

The comparison between $\tau_{\rm He}$ and $\tau_{\rm He,c}$ is known to be ambiguous for two reasons. First, the definition of the acoustic surface 
is uncertain, and is typically defined as a point in the atmosphere where the linear extrapolation of $c^2$ from outermost convective layers 
vanishes \citep{balm90}. This point falls at an acoustic height of approximately 225s from the photosphere for the Sun \citep[see e.g.][]{houd07}. 
Assuming that the ``true" acoustic surface lies between the photosphere and the point where extrapolated $c^2$ vanishes, a maximum systematic 
uncertainty of 225s is expected on $\tau_{\rm He,c}$ for the Sun. Since we only used $\tau_{\rm He,c}$ for sanity check of the fits, we 
simplistically assumed the radial coordinate of the outermost layer in the model at an optical depth of $10^{-5}$ for $R_*$. Secondly, the 
definition of $R_{\rm He}$ in Eq.~\ref{tau} is also uncertain, introducing an additional systematic uncertainty on $\tau_{\rm He,c}$. It has recently 
been shown that the helium glitch signature arises from a region close to the peak in the first adiabatic index, $\Gamma_1$, between the first and 
second helium ionization zones \citep{broo14,verm14b}. We used the radial coordinate of the $\Gamma_1$-peak for $R_{\rm He}$. 

Figure~\ref{fig9} shows the absolute difference between $\tau_{\rm He}$ and $\tau_{\rm He,c}$ for all representative models of all three stars. 
As we can see, the differences are much smaller than the maximum possible systematic uncertainty expected from the uncertainties in the definitions 
of $R_*$ and $R_{\rm He}$, reassuring the quality of the fits and conclusions of the paper.

\section{Discussion and Conclusions}
%-----------------------------------
\label{conc}
We identified three stars, KIC 2837475, 9139163 and 11253226, from the {\it Kepler} asteroseismic LEGACY sample for which the standard models of 
atomic diffusion predict large settling (or complete depletion) of the surface helium. We extracted the oscillatory signature of helium ionization 
from the observed oscillation frequencies. The detection of a strong helium signature in these stars already indicates the presence of significant 
amount of helium in their envelope, in contrast to what we would anticipate if atomic diffusion is the only process determining the surface helium 
abundance. This confirms the presence of additional physical processes competing against atomic diffusion.

In the current study, we assumed turbulence to be the only process that slows down the settling of helium. Since radiative acceleration for
helium is much smaller than gravitational acceleration \citep[see e.g.][]{rich00}, radiative forces have negligible effect on the surface helium 
abundance. The inclusion of radiative forces in the current analysis is only expected to reduce the shift necessary in $[{\rm Fe}/{\rm H}]_i$ to 
compensate for element settling. The advantage of excluding radiative forces is that the detailed grid-based modelling taking uncertainties in 
stellar properties into account becomes computationally feasible.

We implemented in MESA the phenomenological turbulent diffusion coefficient of \citet{rich00} in a form described in \citet{mich11a,mich11b}. In 
this formalism, the surface abundances depend only on one parameter, $\Delta M_0$, which is related to the mass mixed by turbulence in the envelope.
We computed three grids of models with $\Delta M_0 = 5 \times 10^{-6}, 5 \times 10^{-5} \ {\rm and} \ 5 \times 10^{-4}$M$_\odot$ for each star, and 
identified 50 representative models of the star from each grid. Subsequently, we extracted the helium signature from the oscillation frequencies of
representative models, and computed the corresponding average amplitude. For all three stars, models with $\Delta M_0 = 5 \times 10^{-6} \ 
{\rm and} \ 5 \times 10^{-5}$M$_\odot$ have too low average amplitude of helium signature because of the large helium settling to reproduce the 
observed $\langle A_\nu \rangle$. We found that the models with $\Delta M_0 = 5 \times 10^{-4}$M$_\odot$ have average amplitude consistent with the 
observations.

Previous studies using measurements of heavy element abundances and stellar models with atomic diffusion, radiative forces and turbulence
suggested an envelope mixed mass of approximately $10^{-6}$M$_\odot$ \citep[see e.g.][]{mich11b}. This is too small to reproduce the surface helium 
abundance. This clearly demonstrates the potential of using the spectroscopically measured elemental abundances together with the seismic constraint 
of helium abundance to discriminate among the different possible physical processes competing against atomic diffusion. A systematic study in this 
direction using the seismic constraint of helium abundance and the spectroscopic measurements of the heavy element abundances together with the 
state-of-the-art stellar models including atomic diffusion, radiative forces, turbulence, mass loss, etc. can help us identify the most important 
physical processes competing against atomic diffusion, leading to the correct interpretation of the observed surface abundances of hot stars, 
consistent use of atomic diffusion for modelling both hot and cool stars reducing the systematic uncertainties on the inferred stellar properties 
and shed some light on the long-standing cosmological lithium problem.

\section*{Acknowledgements}
Funding for the Stellar Astrophysics Centre is provided by The Danish National Research Foundation (Grant agreement no.: DNRF106). We thank the 
anonymous referee for constructive feedback. We thank J.~Christensen-Dalsgaard and G.~Houdek for carefully reading the manuscript. KV would like to 
thank S.~Hanasoge for kindly providing the time on SEISMO-cluster in the beginning of the project. VSA acknowledges support from the Independent 
Research Fund Denmark (Research grant 7027-00096B).

%\bibliographystyle{/Users/kuldeep/Dropbox/paper/mnras}
%\bibliography{/Users/kuldeep/Dropbox/paper/references}

\bsp
\label{lastpage}
\end{document}